\documentclass[12pt,oneside,reqno,titlepage, hyperfootnotes=false]{amsart}
\usepackage{lmodern}
\usepackage[T1]{fontenc}
\usepackage[latin9]{inputenc}
\usepackage[a4paper]{geometry}
\usepackage{color}
\usepackage{bm}
\usepackage{amstext}
\usepackage{amsthm}
\usepackage{amssymb}
\usepackage{setspace}
\usepackage[authoryear]{natbib}
\usepackage[unicode=true,pdfusetitle,
 bookmarks=true,bookmarksnumbered=false,bookmarksopen=false,
 breaklinks=false,pdfborder={0 0 0},pdfborderstyle={},backref=false,colorlinks=true]
 {hyperref}

\makeatletter
\numberwithin{equation}{section}
\numberwithin{figure}{section}

\usepackage{amsfonts}
\usepackage{amstext}
\usepackage{amsthm}

\usepackage{hyperref}
\hypersetup{
    colorlinks = true,
    citecolor = black
}

\usepackage{threeparttable}
\usepackage{graphicx}
\usepackage{enumerate}
\usepackage{listings}
\usepackage{subcaption}

\usepackage{float}

\newtheorem{thm}{Theorem}

\newtheorem*{lem*}{Lemma}
\newtheorem*{thm*}{Theorem}

\newtheorem*{asm1}{Assumption 1}
\newtheorem*{asm2}{Assumption 2}

\theoremstyle{definition}

\makeatother

\begin{document}
\title{Neyman allocation is minimax optimal for best arm identification with
two arms}
\author{Karun Adusumilli$^\dagger$ }
\begin{abstract}
This note describes the optimal policy rule, according to the local
asymptotic minimax regret criterion, for best arm identification when
there are only two treatments. It is shown that the optimal sampling
rule is the Neyman allocation, which allocates a constant fraction
of units to each treatment in a manner that is proportional to the
standard deviation of the treatment outcomes. When the variances are
equal, the optimal ratio is one-half. This policy is independent of
the data, so there is no adaptation to previous outcomes. At the end
of the experiment, the policy maker adopts the treatment with higher
average outcomes.
\end{abstract}

\thanks{\textit{This version}: \today{}\\
\thispagestyle{empty}I would like to thank Tim Armstrong, Pepe Montiel-Olea
and Chao Qin for helpful comments.\\
$^\dagger$Department of Economics, University of Pennsylvania}
\maketitle

\section{Introduction \label{sec:Introduction}}

The goal of running an experiment is often to determine the best possible
treatment out of a set of candidate treatments. Suppose an agent is
allowed $n$ periods of experimentation. The agent can adaptively
choose which treatment to sample in each period depending on all the
information from the previous periods. This is the Best Arm Identification
(BAI) problem with a fixed budget. The aim is to describe the optimal
sampling and allocation rules for maximizing welfare in the implementation
phase following experimentation. However, in-sample outcomes are not
included in the welfare calculation, thus differentiating this setup
from the multi-armed bandit problem. 

Early on this setting was done by \citet{mannor2004sample}, \citet{chen2000simulation}
and \citet{glynn2004large}. More recently, there has been a surge
of interest in this problem as evidenced by the important work of
\citet{garivier2016optimal}, \citet{russo2016simple}, \citet{carpentier2016tight},
\citet{qin2017improving}, \citet{kasy2019adaptive}, among others.
The seminal analysis of \citet{russo2016simple} characterized the
optimal rate of posterior convergence to the best arm in the fixed
reward gap regime (i.e., when the mean difference in outcomes between
the arms is held fixed, i.e., unchanged over $n$). \citet{carpentier2016tight}
suggest a lower bound on the probability of mis-classification (also
under a fixed gap), and a tight bound for this was later derived by
\citet{kato2022best} in the small gap regime. However these results
do not describe which algorithms to pick under statistical measures
of risk such as minimax regret.\footnote{Recall that regret is defined as the difference in welfare from employing
the treatment chosen by the algorithm as opposed to employing the
best treatment. } 

This paper focuses on the special case of the BAI problem with just
two arms. In this setting, we characterize the optimal policy according
to the local asymptotic minimax regret criterion. The local asymptotic
regime, also known as the diffusion regime, reduces the problem to
the question of choosing the best treatment when the outcomes from
each treatment correspond to a Gaussian process. It is shown that
the optimal sampling rule is the Neyman allocation. It allocates a
constant fraction of units to one of the arms in a manner that is
proportional to the treatment standard deviations, with more variable
treatments being sampled more often. When the treatment variances
are equal, the optimal sampling ratio is one-half. Somewhat surprisingly,
the sampling rule is independent of the data, so there is no adaptation
to previous outcomes. At the end of the experiment, the agent chooses
the treatment with the higher average outcomes. 

The Neyman allocation is well known in the design of experiments literature
as the allocation rule that minimizes estimation variance of the treatment
effect. Despite the difference in goals between estimation and best
arm identification, our results show that the optimal sampling rule
remains unchanged. Thus, as a practical matter, there is no benefit
to running an adaptive experiment (as opposed to a standard RCT) when
there are only two treatments, and the goal is to minimize maximum
regret. 

We emphasize, though, that the above results are intimately connected
to the minimax regret criterion. It is far from obvious that minimax
optimal rule should be non-adaptive. In the Bayesian formulation of
the problem, with normal priors, the optimal sampling rule changes
with time, albeit in a deterministic fashion (\citealp{chick2012sequential,liang2022dynamically}).
With other priors, the sampling rule could be very different. Indeed,
our minimax rule is itself supported by a specific two-point least-favorable
prior. The results thus highlight the important role played by the
prior, and the sensitivity of the optimal decision rules to it. 

Our results are asymptotic in nature, i.e., our policies minimize
maximum regret in the limit as $n\to\infty$. If one instead considers
the global minimax regret criterion (i.e., minimax regret for fixed
$n$), it is known that the sampling ratio $\gamma=1/2$ is optimal
under bounded outcomes; see, \citet{stoye2012new}. In the global
minimax regime, nature chooses the distribution of outcomes for each
treatment, not just the means (so it can choose the variances as well).
Here, it is easy to see that the minimax regret under unbounded rewards
is infinity. The results here are instead derived under local asymptotics.
In the large $n$ limit, previous work by the author, \citet{adusumilli2021risk},
showed that we can treat the outcome distributions as effectively
Gaussian. Furthermore, in this regime, it is without loss of generality
to assume the treatment variances to be known, as replacing the unknown
variances with consistent estimates does not affect asymptotic regret.
The optimal policies under the local asymptotic and global minimax
criteria generally do not coincide, see, e.g., the discussion in \citet{stoye2012new}
and \citet{hirano2009asymptotics}. In the present setting, they coincide
if the treatment variances are equal, but not otherwise.

\section{Setup in the diffusion regime\label{sec:Diffusion-asymptotics-and}}

We start by describing the continuous time version of the problem.
There are two treatments $0,1$ corresponding to mean rewards $\mu_{1},\mu_{0}$
and reward variances $\sigma_{1},\sigma_{0}$. To begin with, we assume
$\sigma_{1},\sigma_{0}$ are known. The experiment runs until time
$t=1$. At each instant, the agent can choose to direct attention
to one of the treatments by choosing the sampling rule $\pi_{1}(\cdot)$.
Here $\pi_{1}(\cdot)$ denotes the probability that treatment 1 is
sampled, and we also define $\pi_{0}=1-\pi_{1}$. At the end of the
experiment, the agent selects a treatment for full-scale implementation,
according to $\pi^{\textrm{fs}}(\cdot)\in\{0,1\}$. 

Let $x_{1}(t),x_{0}(t)$ denote the cumulative outcomes from both
treatments, and $q_{1}(t),q_{0}(t)$ the number of times each treatment
has been sampled so far. Under the policy $\pi_{1}(\cdot)$, these
state variables evolve as diffusion processes (for $a\in\{0,1\}$):
\begin{align}
dx_{a}(t) & =\mu_{a}\pi_{a}dt+\sigma_{a}\sqrt{\pi_{a}}dW_{a}(t)\label{eq:diffiusion evolution 1}\\
dq_{a}(t) & =\pi_{a}dt.\label{eq:diffusion evolution 2}
\end{align}
Here $W_{1}(t),W_{0}(t)$ are Weiner processes tracking the variability
of rewards for each treatment. Since we only sample one treatment
at any given instant, it is without loss of generality to assume $W_{1}(\cdot),W_{0}(\cdot)$
are independent. 

Denote $s(t)=(x_{1}(t),x_{0}(t),q_{1}(t),q_{0}(t))$. We require $\pi_{1}(\cdot)$
to be adapted to the filtration $\{\mathcal{F}_{t}\}_{t\in\mathbb{R}^{+}}$,
where $\mathcal{F}_{t}=\sigma\{s(u);u\le t\}$ is the $\sigma$-algebra
generated by the state variables $s(\cdot)$ until time $t$. Furthermore,
$\pi^{\textrm{fs }}$ needs to be $\mathcal{F}_{t=1}$ measurable.
The aim is choose $\pi_{1},\pi^{\textrm{fs}}$ optimally in order
to minimize maximum regret 
\begin{align*}
R^{\textrm{max}}(\pi_{1},\pi^{\textrm{fs}}) & =\max_{(\mu_{1},\mu_{0})}R\left(\pi_{1},\pi^{\textrm{fs}},(\mu_{1},\mu_{0})\right),\ \textrm{where}\\
R\left(\pi_{1},\pi^{\textrm{fs}},(\mu_{1},\mu_{0})\right) & =\mathbb{E}_{\pi_{1}\vert\mu_{1},\mu_{0}}\left[\max\{\mu_{1}-\mu_{0},0\}-(\mu_{1}-\mu_{0})\pi^{\textrm{fs}}\right],
\end{align*}
where $\mathbb{E}_{\pi_{1}\vert\mu_{1},\mu_{0}}[\cdot]$ denotes the
expectation under the sampling rule $\pi_{1}$ given $\mu_{1},\mu_{0}$. 

\section{Minimax optimal policy in the diffusion regime}

Following \citet{wald1945statistical}, we model the minimax problem
as a game played between nature and the agent. Nature chooses a prior
$m_{0}$ over the vector of mean rewards $(\mu_{1},\mu_{0})$ and
the agent chooses the policy functions $(\pi_{1},\pi^{\textrm{fs}})$.
The equilibrium value of this game gives minimax regret. The optimal
choice of the agent is termed the minimax policy, and that of nature,
the least favorable prior. 

In general, solving the two-player game can be daunting. However,
a useful starting point is the conjecture, made in \citet{adusumilli2021risk},
that the least favorable has the same number of support points as
the number of treatments, which is 2 in this setting (we will formally
verify this conjecture here). 

Following the above ansatz, consider a prior supported on the two
points $(a_{1},b_{1}),(a_{0},b_{0})$, where $a_{1}>b_{1}$ and $b_{0}>a_{0}$.
Let $\theta=1$ denote the state when nature chooses $(a_{1},b_{1})$,
and $\theta=0$ the state when nature chooses $(a_{0},b_{0})$. Also
let $(\Omega,\mathbb{P},\mathcal{F}_{t})$ denote the relevant probability
space, where $\mathcal{F}_{t}$ is defined above. Define the probability
measures $P_{0},P_{1}$ as $P^{0}:=\mathbb{P}(A\vert\theta=0)$ and
$P^{1}:=\mathbb{P}(A\vert\theta=1)$ for any $A\in\mathcal{F}_{t}$.

Noting that $W_{1}(\cdot),W_{0}(\cdot)$ are independent of each other,
it follows by the Girsanov theorem (see also \citealp[Section 4.2.1]{shiryaev2007optimal})
that
\begin{align}
\ln\frac{dP^{1}}{dP^{0}}(\mathcal{F}_{t}) & =\frac{(a_{1}-a_{0})}{\sigma_{1}^{2}}x_{1}(t)+\frac{(b_{1}-b_{0})}{\sigma_{0}^{2}}x_{0}(t)-\frac{(a_{1}^{2}-a_{0}^{2})}{2\sigma_{1}^{2}}q_{1}(t)-\frac{(b_{1}^{2}-b_{0}^{2})}{2\sigma_{0}^{2}}q_{0}(t).\label{eq:LR process}
\end{align}
Let $m_{1}$ denote the prior probability that $\theta=1$. Furthermore,
let $m_{t}^{\pi}=\mathbb{P}(\theta=1\vert\mathcal{F}_{t})$ denote
the posterior probability that $\theta=1$. Following \citet[Section 4.2.1]{shiryaev2007optimal},
the belief process $m_{t}^{\pi}$ can be related to the likelihood
ratio process $\varphi(t):=\frac{dP^{1}}{dP^{0}}(\mathcal{F}_{t})$
as 
\[
m_{t}^{\pi}=\frac{m_{1}\varphi(t)}{(1-m_{1})+m_{1}\varphi(t)}.
\]
The optimal allocation rule at the end of the experiment is then 
\begin{align}
\pi^{*\textrm{fs}} & =\mathbb{I}\left\{ a_{1}m_{1}^{\pi}+a_{0}(1-m_{1}^{\pi})\ge b_{1}m_{1}^{\pi}+b_{0}(1-m_{1}^{\pi})\right\} \nonumber \\
 & =\mathbb{I}\left\{ \ln\varphi(t)\ge\ln\frac{(b_{0}-a_{0})(1-m_{1})}{(a_{1}-b_{1})m_{1}}\right\} \label{eq:allocation rule}
\end{align}

We can obtain the Nash equilibrium of this game under the following
steps:

\subsection*{1. Indifference priors.}

Consider two point priors supported on\\
 $(\sigma_{1}\Delta/2,-\sigma_{0}\Delta/2),(-\sigma_{1}\Delta/2,\sigma_{0}\Delta/2$)
for some $\Delta>0$. For these priors, the agent is indifferent between
the choice of any measurable $\pi_{1}$. Intuitively, either treatment
is equally informative about the true value of $\theta\in\{0,1\}$
here, so it does not matter which treatment the agent directs attention
to. 

To see this formally, observe that under these priors, (\ref{eq:LR process})
implies 
\begin{equation}
\ln\varphi(t):=\ln\frac{dP^{1}}{dP^{0}}(\mathcal{F}_{t})=\Delta\left\{ \frac{x_{1}(t)}{\sigma_{1}}-\frac{x_{0}(t)}{\sigma_{0}}\right\} .\label{eq:LR process indiffierence}
\end{equation}
Suppose $\theta=1$. Then the evolution equations (\ref{eq:diffiusion evolution 1}),
(\ref{eq:diffusion evolution 2}) imply 
\begin{align*}
\frac{dx_{1}(t)}{\sigma_{1}}-\frac{dx_{0}(t)}{\sigma_{0}} & =\frac{\Delta}{2}dt+\sqrt{\pi_{1}}dW_{1}(t)-\sqrt{\pi_{0}}dW_{0}(t)\\
 & =\frac{\Delta}{2}dt+d\tilde{W}(t),
\end{align*}
where $\tilde{W}(t):=\sqrt{\pi_{1}}dW_{1}(t)-\sqrt{\pi_{0}}dW_{0}(t)$
is a one dimensional Weiner process, being a linear combination of
two independent Weiner processes with $\pi_{1}+\pi_{0}=1$. Plugging
the above into (\ref{eq:LR process indiffierence}) gives 
\[
d\ln\varphi(t)=\frac{\Delta^{2}}{2}dt+\Delta d\tilde{W}(t).
\]
In a similar manner, we can show under $\theta=0$ that 
\[
d\ln\varphi(t)=-\frac{\Delta^{2}}{2}dt+\Delta d\tilde{W}(t).
\]
In either case, the choice of $\pi_{1}$ does not affect the evolution
of the likelihood-ratio process $\varphi(t)$, and consequently has
no bearing on the evolution of the beliefs $m_{t}^{\pi}$. 

\subsection*{2. Form of $\pi^{\textrm{fs}}$ under the indifference prior.}

Suppose nature chooses the indifference prior for some $\Delta>0$.
Then (\ref{eq:allocation rule}) and (\ref{eq:LR process indiffierence})
imply 
\begin{align*}
\pi^{*\textrm{fs}} & =\mathbb{I}\left\{ \Delta\left\{ \frac{x_{1}(1)}{\sigma_{1}}-\frac{x_{0}(1)}{\sigma_{0}}\right\} \ge\ln\frac{(1-m_{1})}{m_{1}}\right\} \\
 & =\mathbb{I}\left\{ \frac{x_{1}(1)}{\sigma_{1}}-\frac{x_{0}(1)}{\sigma_{0}}\ge c\right\} ,
\end{align*}
where $c:=\Delta^{-1}\ln\frac{(1-m_{1})}{m_{1}}$. Note that for any
given $\Delta$ we can induce $c$ to be any value we like by varying
$m_{1}$. This implies that the choice of $c$ is essentially equivalent
to the choice of $m_{1}$. 

\subsection*{3. Determining the equilibrium values of $\pi_{1},c$.}

Based on the above observations, consider the ansatz that the minimax
policy of the agent is of the form $\pi_{1}=\gamma$ for some $\gamma\in[0,\infty)$
and 
\[
\tilde{\pi}^{\textrm{fs}}=\mathbb{I}\left\{ \frac{x_{1}(1)}{\sigma_{1}}-\frac{x_{0}(1)}{\sigma_{0}}\ge c\right\} .
\]

Suppose nature chooses $(\tilde{a}_{1},\tilde{b}_{1})$, where $\tilde{a}_{1}>\tilde{b}_{1}$.
Under the ansatz for the minimax policy,
\[
\frac{dx_{1}(t)}{\sigma_{1}}-\frac{dx_{0}(t)}{\sigma_{0}}=\left(\tilde{a}_{1}\frac{\gamma}{\sigma_{1}}-\tilde{b}_{1}\frac{1-\gamma}{\sigma_{0}}\right)dt+d\tilde{W}(t),
\]
so the expected regret under a given sampling probability $\gamma$,
and threshold constant $c$ is 
\begin{align}
R\left(\gamma,c,(\tilde{a}_{1},\tilde{b}_{1})\right) & =(\tilde{a}_{1}-\tilde{b}_{1})\mathbb{P}\left(\frac{x_{1}(1)}{\sigma_{1}}-\frac{x_{0}(1)}{\sigma_{0}}\le c\right)\nonumber \\
 & =(\tilde{a}_{1}-\tilde{b}_{1})\mathbb{P}\left(\left\{ \tilde{a}_{1}\frac{\gamma}{\sigma_{1}}-\tilde{b}_{1}\frac{1-\gamma}{\sigma_{0}}\right\} +\tilde{W}(1)\le c\right)\nonumber \\
 & =(\tilde{a}_{1}-\tilde{b}_{1})\Phi\left(c-\left\{ \tilde{a}_{1}\frac{\gamma}{\sigma_{1}}-\tilde{b}_{1}\frac{1-\gamma}{\sigma_{0}}\right\} \right).\label{eq:nature BR 1}
\end{align}
If $\gamma/\sigma_{1}>(1-\gamma)/\sigma_{0}$, nature can choose $\tilde{a}_{1},\tilde{b}_{1}$
in such a way that the expected regret above is arbitrarily large
(nature can set $\tilde{a}_{1}-\tilde{b}_{1}$ to be arbitrarily large
and $\tilde{a}_{1}$ to be negative with $\vert\tilde{a}_{1}\vert\gg\tilde{a}_{1}-\tilde{b}_{1}$;
the latter ensures the term inside the $\Phi(\cdot)$ function in
(\ref{eq:nature BR 1}) is close to $\infty$). This suggests that
the optimal sampling rule should satisfy $\gamma^{*}\le\sigma_{1}/(\sigma_{1}+\sigma_{0})$. 

Similarly, if nature chooses $(\tilde{a}_{0},\tilde{b}_{0})$ where
$\tilde{b}_{0}>\tilde{a}_{0}$, the expected regret is 
\begin{equation}
R\left(\gamma,c,(\tilde{a}_{0},\tilde{b}_{0})\right)=(\tilde{b}_{0}-\tilde{a}_{0})\Phi\left(-c-\left\{ \tilde{b}_{0}\frac{1-\gamma}{\sigma_{0}}-\tilde{a}_{0}\frac{\gamma}{\sigma_{1}}\right\} \right).\label{eq:nature BR 2}
\end{equation}
Now if $\gamma/\sigma_{1}<(1-\gamma)/\sigma_{0}$, nature can again
achieve infinite regret in (\ref{eq:nature BR 2}). Combined with
the previous observation that $\gamma^{*}\le\sigma_{1}/(\sigma_{1}+\sigma_{0})$,
we are led to conclude only $\gamma^{*}=\sigma_{1}/(\sigma_{1}+\sigma_{0})$
can prevent infinite regret in both scenarios. So setting $\gamma$
to this value and combining (\ref{eq:nature BR 1}) - (\ref{eq:nature BR 2}),
we find the max regret under $(\gamma^{*},c)$ to be 
\begin{equation}
R^{\max}(\gamma^{*},c)=\max\left\{ \max_{\delta}\delta\Phi\left(c-\frac{\delta}{\sigma_{1}+\sigma_{0}}\right),\max_{\delta}\delta\Phi\left(-c-\frac{\delta}{\sigma_{1}+\sigma_{0}}\right)\right\} .\label{eq:max regret - gamma, c}
\end{equation}
Clearly, $R^{\max}(\gamma^{*},c)$ is minimized when $c^{*}=0$. 

We thus find that the optimal choices of $\gamma,c$ for the agent
are $\sigma_{1}/(\sigma_{1}+\sigma_{0}),0$. It remains to show that
this constitutes an equilibrium. To this end, suppose that the agent
chooses $\gamma^{*}=\sigma_{1}/(\sigma_{1}+\sigma_{0}),c^{*}=0$.
Then it follows from (\ref{eq:nature BR 1}) - (\ref{eq:max regret - gamma, c})
that the optimal response of nature is to choose $(\tilde{a},\tilde{b})$
such that
\begin{align*}
\left|\tilde{a}-\tilde{b}\right|=\eta^{*} & :=\arg\max_{\delta}\delta\Phi\left(-\frac{\delta}{\sigma_{1}+\sigma_{0}}\right)\\
 & =(\sigma_{1}+\sigma_{0})\arg\max_{\delta}\delta\Phi\left(-\delta\right).
\end{align*}
Nature is otherwise indifferent between any $(\tilde{a},\tilde{b})$
satisfying the above condition. In particular, a prior supported on
the two points $(\sigma_{1}\Delta^{*}/2,-\sigma_{0}\Delta^{*}/2),(-\sigma_{1}\Delta^{*}/2,\sigma_{0}\Delta^{*}/2$),
where $\Delta^{*}:=2\eta^{*}/(\sigma_{1}+\sigma_{0})=2\arg\max_{\delta}\delta\Phi\left(-\delta\right)$,
would be a best response to the agent's actions. This is an indifference
prior, so the agent's actions are best responses to it as well: the
agent is indifferent between any $\pi_{1}$ as noted earlier, and
we can induce $c^{*}=0$ by setting $m_{1}=1/2$. We have thus obtained
a Nash equilibrium to the game. The result is summarized below:

\begin{thm} \label{Thm: Minimax policy} The minimax optimal decision
rule is $\bm{d}^{*}:=(\pi_{1}^{*},\pi^{*\textrm{fs}})$, where $\pi_{a}^{*}=\sigma_{a}/(\sigma_{1}+\sigma_{0})$
for $a\in\{0,1\}$, and $\pi^{*\textrm{fs}}=\mathbb{I}\left\{ \frac{x_{1}(1)}{\sigma_{1}}-\frac{x_{0}(1)}{\sigma_{0}}\ge0\right\} $.
Furthermore, the least favorable prior is a symmetric two-point distribution
supported on $(\sigma_{1}\Delta^{*}/2,-\sigma_{0}\Delta^{*}/2)$ and
$(-\sigma_{1}\Delta^{*}/2,\sigma_{0}\Delta^{*}/2)$ where $\Delta^{*}=2\arg\max_{\delta}\delta\Phi\left(-\delta\right)$.
\end{thm}

\subsection*{Discussion of the minimax policy}

Perhaps the most striking feature of the minimax optimal policy is
that it is independent of the data. The policy assigns a fixed proportion
$\gamma=\sigma_{1}/(\sigma_{1}+\sigma_{0})$ of units to treatment
1. This is just the Neyman allocation. It is same sampling rule one
would also employ if the aim were to minimize the estimation variance
of the treatment effect $\mu_{1}-\mu_{0}$. 

The implementation rule at the end of the experiment is given by 
\[
\pi^{*\textrm{fs}}=\mathbb{I}\left\{ \frac{x_{1}(1)}{\sigma_{1}}-\frac{x_{0}(1)}{\sigma_{0}}\ge0\right\} .
\]
Under the optimal sampling rule, treatment 1 is sampled $\sigma_{1}/\sigma_{0}$
times more often than treatment 0. This implies $\frac{x_{1}(1)}{\sigma_{1}}-\frac{x_{0}(1)}{\sigma_{0}}$
is proportional to the difference in average outcomes between both
treatments. In other words, the treatment with higher average outcomes
is chosen for full-scale implementation. 

\section{Local asymptotics}

So far the optimal policies have been characterized under the diffusion
regime. Let $V^{*}$ denote the value of maximum regret from the previous
section
\begin{align*}
V^{*} & :=\max_{\delta}\delta\Phi\left(-\frac{\delta}{\sigma_{1}+\sigma_{0}}\right)=(\sigma_{1}+\sigma_{0})\max_{\delta}\delta\Phi\left(-\delta\right).
\end{align*}
Then $V^{*}$ is also the lower bound on asymptotic minimax regret
in both parametric and non-parametric regimes. In fact the bound is
tight in the sense that it can be achieved through analogues of the
minimax optimal policy described above. Formal statements follow:

\subsection{Parametric regime}

Let $\left\{ \left(P_{\theta^{(1)}}^{(1)},P_{\theta^{(0)}}^{(0)}\right):\theta^{(1)},\theta^{(0)}\in\mathbb{R}\right\} $
denote the set of candidate probability measures for the joint distribution
of outcomes under both treatments. It is without loss of generality
to assume $P_{\theta^{(1)}}^{(1)},P_{\theta^{(0)}}^{(0)}$ are mutually
independent (conditional on $\theta^{(1)},\theta^{(0)}$) as we only
ever observe the outcomes from one treatment anyway. 

The mean outcomes under a parameter $\theta$ are given by $\mu_{a}(\theta):=\mathbb{E}_{P_{\theta}^{(a)}}[Y_{i}]$.
Following \citet{hirano2009asymptotics} and \citet{adusumilli2021risk},
for $a\in\{0,1\}$, we consider local perturbations of the form $\{\theta_{0}^{(a)}+h^{(a)}/\sqrt{n};h^{(a)}\in\mathbb{R}\}$
around a reference parameter $\theta_{0}^{(a)}$. As in those papers,
$\theta_{0}^{(a)}$ is chosen such that $\mu_{a}(\theta_{0}^{(a)})=0$
for each $a\in\{0,1\}$. This defines the hardest instance of the
best arm identification problem, with $\mu_{a}(\theta_{0}^{(a)}+h/\sqrt{n})\approx\dot{\mu}_{a}^{\intercal}h/\sqrt{n}$
where $\dot{\mu}_{a}:=\nabla_{\theta}\mu_{a}(\theta_{0})$. Denote
$P_{h}^{(a)}:=P_{\theta_{0}^{(a)}+h/\sqrt{n}}^{(a)}$ and let $\mathbb{E}_{h}^{(a)}[\cdot]$
denote its corresponding expectation. We assume $P_{\theta}^{(a)}$
is differentiable in quadratic mean around $\theta_{0}^{(a)}$ with
score functions $\psi_{a}(Y_{i})$ and information matrices $I_{a}:=\left(\mathbb{E}_{0}^{(a)}[\psi^{2}]\right)^{-1}$.
Let $\hat{\mu}_{n,a}$ denote the best regular estimator of the mean
outcomes for treatment $a$ in the sense
\begin{equation}
\sqrt{n}\left(\hat{\mu}_{n,a}-\mu_{a}(h)\right)\xrightarrow[P_{n,h}^{(a)}]{d}\mathcal{N}(0,\dot{\mu}_{a}^{\intercal}I_{a}^{-1}\dot{\mu}_{a}).\label{eq:best regular}
\end{equation}
In what follows, we define $\sigma_{a}^{2}:=\dot{\mu}_{a}^{\intercal}I_{a}^{-1}\dot{\mu}_{a}$. 

Suppose $V_{n}(\pi_{1},\pi^{\textrm{fs}};h_{1},h_{0})$ denotes the
frequentist regret due to the sampling rule $\pi_{1}$ and implementation
rule $\pi^{\textrm{fs}}$ when the local parameters are $\theta_{0}^{(1)}+h_{1}/\sqrt{n}$
and $\theta_{0}^{(0)}+h_{0}/\sqrt{n}$. Define $n_{a}:=n\sigma_{a}/(\sigma_{1}+\sigma_{0});\ a\in\{0,1\}$.
Our key result in this section is that the policy rules $\pi_{1}^{*},\pi^{\textrm{fs}}$,
given below are minimax optimal:
\begin{align*}
\pi_{1}^{*} & =\frac{\sigma_{1}}{\sigma_{1}+\sigma_{0}};\quad\pi^{*\textrm{fs}}=\mathbb{I}\left\{ \hat{\mu}_{n_{1},1}-\hat{\mu}_{n_{0},0}\ge0\right\} .
\end{align*}
This is shown under the following assumptions.

\begin{asm1} (i) The class $\{P_{\theta}^{(a)};\theta\in\mathbb{R}\}$
is differentiable in quadratic mean around $\theta_{0}^{(a)}$ for
$a\in\{0,1\}$. 

(ii) $\mathbb{E}_{0}^{(a)}[\exp\vert\psi_{a}(Y)\vert]<\infty$ for
$a\in\{0,1\}$. 

(iii) For each $a\in\{0,1\}$ there exists $\vert\dot{\mu}_{a}\vert<\infty$
s.t $\sqrt{n}\mu\left(P_{h}^{(a)}\right)=\dot{\mu}_{a}^{\intercal}h+o(\vert h\vert^{2})$.

(iv) The estimators $\hat{\mu}_{a}$ are best regular as defined in
(\ref{eq:best regular}). \end{asm1}

\begin{thm} \label{Thm: General parametric families}Suppose that
Assumptions 1(i)-(iii) hold. Then:

(i) $\lim_{n\to\infty}\inf_{\pi_{1},\pi^{\textrm{fs}}}\sup_{\vert h_{1}\vert,\vert h_{0}\vert\le C}V_{n}(\pi_{1},\pi^{\textrm{fs}};h_{1},h_{0})\ge V^{*}$
for any $C<\infty$. 

(ii) If, further, Assumption 1(iv) holds, 
\[
\sup_{\mathcal{J}}\lim_{n\to\infty}\sup_{\vert h_{1}\vert,\vert h_{0}\vert\in\mathcal{J}}V_{n}(\pi_{1}^{*},\pi^{*\textrm{fs}};h_{1},h_{0})=V^{*},
\]
where the outer supremum is taken over all finite subsets $\mathcal{J}$
of $\mathbb{R}$. \end{thm}

The first part of Theorem 1 says that $V^{*}$ provides a lower bound
on minimax regret. This is shown in \citet{adusumilli2021risk}. Strictly
speaking, the results in \citet{adusumilli2021risk} characterize
the minimax regret using PDE methods. However, from the results in
Section 3, it is straightforward to see that the regret can be alternatively
characterized using $V^{*}$ defined above. 

The second part of the theorem says that $\pi_{1}^{*},\pi^{\textrm{fs}}$
attain this bound, thereby proving that they are asymptotically minimax
optimal. The proof of this uses similar arguments in \citet[Lemma 3 \& Theorem 3.2]{hirano2009asymptotics},
and is therefore omitted. 

\subsection{Non-parametric regime}

Let $\mathcal{P}_{1},\mathcal{P}_{0}$ denote a candidate class of
probability measures for the two treatments with bounded variances,
and dominated by some measure $\nu$. Also, let $P_{0}^{(1)}\in\mathcal{P}_{1}$
and $P_{0}^{(0)}\in\mathcal{P}_{0}$ denote reference probability
distributions. Following \citet{van2000asymptotic}, for each treatment
$a\in\{0,1\}$ we consider smooth one-dimensional sub-models of the
form $\{P_{t,h}^{(a)}:t\le\eta\}$ for some $\eta>0$, where $h(\cdot)$
is a measurable function satisfying 
\begin{equation}
\int\left[\frac{\left(dP_{t,h}^{(a)}\right)^{1/2}-\left(dP_{0}^{(a)}\right)^{1/2}}{t}-\frac{1}{2}h\left(dP_{0}^{(a)}\right)^{1/2}\right]^{2}d\nu\to0\ \textrm{as}\ t\to0.\label{eq:qmd non-parametrics}
\end{equation}
In analogy with the parametric setting, we compute minimax regret
under the local (i.e., local to $P_{0}^{(1)},P_{0}^{(0)}$) sequence
of probability measures $P_{1/\sqrt{n},h}^{(a)}$. 

It is well known, see e.g \citet{van2000asymptotic}, that (\ref{eq:qmd non-parametrics})
implies $\int hdP_{0}^{(a)}=0$ and $\int h^{2}dP_{0}^{(a)}<\infty$.
The set of all such candidate $h$ is termed the tangent space $T(P_{0}^{(a)})$.
This is a subset of the Hilbert space $L^{2}(P_{0}^{(a)})$, endowed
with the inner product $\left\langle f,g\right\rangle _{a}=\mathbb{E}_{P_{0}^{(a)}}[fg]$
and norm $\left\Vert f\right\Vert _{a}=\left\langle f,f\right\rangle _{a}$.
The mean rewards under $P\in\mathcal{P}_{a}$ are given by $\mu_{a}(P)=\int xdP(x)$.
To obtain non-trivial regret bounds, we suppose $\mu_{a}(P_{0}^{(a)})=0$
for $a\in\{0,1\}$. The rationale for this is similar to setting $\mu_{a}(\theta_{0}^{(a)})=0$
in the parametric setting. Let $\psi_{a}(x):=x-\int xdP_{0}^{(a)}(x)=x$
and $\sigma_{a}^{2}:=\int x^{2}dP_{0}^{(a)}(x)$. Then, $\psi_{a}(\cdot)$
is the influence function corresponding to $\mu_{a}$, in the sense
that under some mild assumptions on $\{P_{t,h}^{(a)}\}$, 
\begin{equation}
\frac{\mu(P_{t,h}^{(a)})-\mu(P_{0}^{(a)})}{t}-\left\langle \psi_{a},h\right\rangle =\frac{\mu(P_{t,h}^{(a)})}{t}-\left\langle \psi_{a},h\right\rangle =o(t).\label{eq:influence function}
\end{equation}

It will be shown that the following policy rules are minimax optimal:
\[
\pi_{1}^{*}=\frac{\sigma_{1}}{\sigma_{1}+\sigma_{0}};\quad\pi^{*\textrm{fs}}=\mathbb{I}\left\{ \frac{x_{1}(1)}{\sigma_{1}}-\frac{x_{0}(1)}{\sigma_{0}}\ge0\right\} ,
\]
where $x_{a}(t):=n^{-1/2}\sum_{i=1}^{\left\lfloor nt\right\rfloor }\mathbb{I}\{A_{i}=a\}Y_{i}$
with $A_{i}\in\{0,1\}$ indicating which treatment was sampled in
period $i$. The following assumptions are taken from \citet{adusumilli2021risk}:

\begin{asm2} (i) The sub-models $\{P_{t,h}^{(a)};h\in T(P_{0})\}$
satisfy (\ref{eq:qmd non-parametrics}). 

(ii) $\mathbb{E}_{P_{0}^{(a)}}[\exp\vert Y\vert]<\infty$ for $a\in\{0,1\}$. 

(iii) For each $a\in\{0,1\}$, $\sqrt{n}\mu(P_{1/\sqrt{n},h}^{(a)})=\left\langle \psi_{a},h\right\rangle +o\left(\left\Vert h\right\Vert ^{2}\right)$.
\end{asm2}

For the statement of the theorem below, let $\mathcal{H}_{I}^{(a)}$
denote a finite subset of the tangent space $T(P_{0}^{(a)})$ of dimension
$I$; this is the space spanned by $I$ sub-elements from an orthonormal
basis $\{\phi_{1},\phi_{2},\dots\}$ for the closure of $T(P_{0}^{(a)})$. 

\begin{thm} \label{non-parametric theorem}Suppose that Assumptions
2(i)-(iii) hold. Then,
\[
\sup_{I_{1},I_{0}}\lim_{n\to\infty}\inf_{\pi\in\Pi}\sup_{h_{1}\in\mathcal{H}_{I_{1}}^{(1)},h_{0}\in\mathcal{H}_{I_{0}}^{(0)}}V_{n}(\pi_{1},\pi^{\textrm{fs}};h_{1},h_{0})\ge V^{*}
\]
where the outer supremum is taken over all finite subsets $I_{1},I_{0}$
of the tangent spaces $T(P_{0}^{(1)}),T(P_{0}^{(0)})$. Furthermore,
\[
\sup_{I_{1},I_{0}}\lim_{n\to\infty}\sup_{h_{1}\in\mathcal{H}_{I_{1}}^{(1)},h_{0}\in\mathcal{H}_{I_{0}}^{(0)}}V_{n}(\pi_{1}^{*},\pi^{\textrm{*fs}};h_{1},h_{0})=V^{*}.
\]
 \end{thm}

The first part of the above theorem shows that the minimax regret
is lower bounded by $V^{*}$. This is already shown in \citet{adusumilli2021risk}.
The definition of minimax regret here is the same as in \citet[Theorem 25.21]{van2000asymptotic}.
The second part of the theorem shows that $\pi_{1}^{*},\pi^{*\textrm{fs}}$
attain this bound and are therefore asymptotically minimax optimal.
The proof of this is similar to \citet[Lemma 3']{hirano2009asymptotics},
and is omitted. 

\subsection{Unknown variances}

Replacing $\sigma_{1},\sigma_{0}$ with consistent estimates has no
effect on asymptotic risk. Since the optimal allocation rule depends
on these quantities, we can proceed as follows: Let $\bar{n}:=n^{\rho}$
for some $\rho\in(0,1)$. For the first $\bar{n}$ observations, we
sample the treatments in equal proportions and use the outcomes generated
to obtain consistent estimates, $\hat{\sigma}_{1},\hat{\sigma}_{0}$
of $\sigma_{1},\sigma_{0}$ (alternatively, one could obtain these
estimates from a pilot experiment). We then apply the Neyman allocation
rule $\hat{\sigma}_{a}/(\hat{\sigma}_{1}+\hat{\sigma}_{0})$ for all
the observations from $\bar{n}$ onwards. The resulting policies also
attain the lower bounds on asymptotic minimax regret, in both parametric
and non-parametric settings. In practice, the choice of $\rho$ would
matter, and further work is needed to determine this. 

\section{Conclusion}

This paper describes the asymptotic minimax optimal policy for best
arm identification with two arms. The optimal sampling rule involves
sampling in fixed proportions and there is no adaptation to past outcomes.
In this setting, the goals of estimation and minimizing regret (at
least according to the minimax criterion) coincide. Crucial to this
result is the observation that nature's least favorable prior has
a two-point support and makes the agent indifferent between any sampling
rule. Going beyond two arms, we expect the least favorable prior to
generally have as many support points as the number of arms. It is
unknown, however, if there exist indifference inducing priors beyond
the two arm case. 

\bibliographystyle{IEEEtranSN}
\bibliography{BAI}

\end{document}